\documentclass{svmult-turin}

\usepackage{type1cm} 

\usepackage{makeidx}         
\usepackage{graphicx}        

\usepackage[bottom]{footmisc}

\usepackage{newtxtext}      %
\usepackage{newtxmath}      

\usepackage{combelow,enumitem}
\usepackage{mathrsfs,amsfonts,amsopn}

\newcommand{\STbR}{\mathbb{R}}
\newcommand{\STbC}{\mathbb{C}}
\newcommand{\STbN}{\mathbb{N}}
\newcommand{\STbZ}{\mathbb{Z}}

\newcommand{\STcA}{\mathcal{A}}
\newcommand{\STcF}{\mathcal{F}}
\newcommand{\STcS}{\mathcal{S}}
\newcommand{\STcG}{\mathcal{G}}
\newcommand{\STcB}{\mathcal{B}}
\newcommand{\STcM}{\mathcal{M}}
\newcommand{\STcH}{\mathcal{H}}
\newcommand{\STcD}{\mathcal{D}}
\newcommand{\STcV}{\mathcal{V}}
\newcommand{\STrd}{\STbR^d}

\newcommand{\STzd}{\STbZ^d}
\newcommand{\STrdd}{\STbR^{2d}}
\newcommand{\STtlpk}{\tilde{L}^p_k}
\newcommand{\SThbi}{\frac{i}{\hbar}}
\newcommand{\STten}{\widetilde{E}^{(N)}}
 
\begin{document}

\title*{A time-frequency analysis perspective on Feynman path integrals}

\author{S. Ivan Trapasso}
\institute{Dipartimento di Scienze Matematiche (DISMA) ``G. L. Lagrange'', Politecnico di Torino. \\ Corso Duca degli Abruzzi 24, 10129 Torino, Italy. \\ \email{salvatore.trapasso@polito.it}}
%
%
\maketitle

\abstract{The purpose of this expository paper is to highlight the starring role of time-frequency analysis techniques in some recent contributions concerning the mathematical theory of Feynman path integrals. We hope to draw the interest of mathematicians working in time-frequency analysis on this topic, as well as to illustrate the benefits of this fruitful interplay for people working on path integrals.}


\section{Introduction}
The path integral formulation of non-relativistic quantum mechanics is a paramount contribution by Richard Feynman (Nobel Prize in Physics, 1965) to modern theoretical physics. The origin of this approach goes back to Feynman's Ph.D. thesis of 1942 at Princeton University (recently reprinted, cf. \cite{STfeyn thesis}) but was first published in the form of research paper in 1948 \cite{STfeyn1 48}; see also \cite{STsauer} for some historical hints. In rough terms we could say that this approach provides a quantum counterpart to Lagrangian mechanics, while the standard framework for canonical quantization as developed by Dirac relies on the Hamiltonian formulation of classical mechanics. Path integrals and Feynman's deep physical intuition were the main ingredients of the celebrated diagrams, introduced in the 1949 paper \cite{STfeyn 49}, which gave a whole new outlook on quantum field theory. 

For a first-hand pedagogical introduction we recommend the textbook \cite{STfeyn2 hibbs}, where it is clarified how the physical intuition of path integrals comes from a deep understanding of the lesson given by the two-slit experiment. We briefly outline below the main features of Feynman's approach. Recall that the state of a non-relativistic particle in the Euclidean space $\STrd$ at time $t$ is represented by the wave function $\psi(t,x)$, $(t,x) \in \STbR \times \STrd$, such that $\psi (t,\cdot)\in L^{2}(\STrd)$. The time-evolution of a state $\varphi(x)$ at $t=0$ is governed by the Cauchy problem for the Schr\"odinger equation:
\begin{equation}
\begin{cases}
i \hbar \partial_{t}\psi=(H_{0}+V(x))\psi\\
\psi(0,x)=\varphi(x),
\end{cases}\label{STcauchy schr}
\end{equation}
where $0<\hbar\le 1$ is a parameter (the Planck constant), $H_{0}=-\hbar^2\triangle/2$ is the standard Hamiltonian for a free particle and $V$ is a real-valued potential; we conveniently set $m=1$ for the mass of the particle. 
The map $U(t,s): \psi(s,\cdot) \mapsto \psi(t,\cdot)$, $t,s\in\STbR$, is a unitary operator on $L^2(\STrd)$ and is known as \textit{propagator}\footnote{We remark that in physics literature the term ``propagator'' is usually reserved to the integral kernel $u_t$ of $U(t)$, see below. This may possibly lead to confusion since it is in conflict with the traditional nomenclature adopted in the analysis of PDEs.} or \textit{evolution operator}; we set $U(t)$ for $U(t,0)$. Since $U(t)$ is a linear operator we can formally represent it as an integral operator, namely
\[
\psi(t,x)=\int_{\STrd}u_t(x,y)\varphi(y)dy,
\]
where the kernel $u_t(x,y)$ (we also write $u_{t,s}(x,y)$ or $u(t,s)(x,y)$ for the kernel of $U(t,s)$) is interpreted as the transition amplitude from the position $y$ at time $0$ to the position $x$ at time $t$. In a nutshell, Feynman's prescription is a recipe for this kernel, the main ingredients being all the possible paths from $y$ to $x$ that the particle could follow. The contribution of each interfering alternative path to the total probability amplitude is a phase factor involving the \emph{action functional} evaluated on the path, that is
\[
S\left[\gamma\right] = S(t,0,x,y)=\int_{s}^{t}L(\gamma(\tau),\dot{\gamma}(\tau))d\tau,
\]
where $L$ is the Lagrangian functional of the underlying classical system. Therefore, the kernel should be formally represented as 
\begin{equation}
u_t(x,y)=\int e^{\SThbi S\left[\gamma\right]}\mathcal{D}\gamma,\label{STker path int}
\end{equation} 
that is a sort of integral over the infinite-dimensional space of paths satisfying the conditions above. This intriguing picture is further reinforced by the following remark: a formal application of the stationary phase method shows that the semiclassical limit $\hbar \to 0$ selects the classical trajectory, hence we recover the principle of stationary action of classical mechanics.

It is well known after Cameron \cite{STcameron} that $\STcD \gamma$ cannot be a Lebesgue-type measure on the space of paths, neither it can be constructed as a Wiener measure with complex variance - it would have infinite total variation. The literature concerning the problem of putting formula \eqref{STker path int} on firm mathematical ground is huge; the interested reader could benefit from the monographs \cite{STalbeverio book,STfujiwara5 book, STmazzucchi} as points of departure. We will describe below only two of the several schemes which have been manufactured in order to give a rigorous meaning to \eqref{STker path int}, the type of techniques involved ranging from geometric to stochastic analysis; these approaches both rely on operator-theoretic strategies and are called \textit{sequential approach} and \textit{time slicing approach}. Basically, one is lead to study sequences of operators on $L^2(\STrd)$ which converge to the exact propagator $U(t)$ in a sense to be specified, the strength of convergence competing against the regularity of the potential $V$. 

This is the point where time-frequency analysis enters the scene. Techniques of phase space analysis are indeed very well suited to the study of path integrals, the reasons being manifold. First of all, pseudodifferential and Fourier integral operators can be effectively treated from a time-frequency analysis perspective as evidenced by a now vast literature - we highlight \cite{STCGNR fio,STCGNR jmp,STgro ped,STgro2 sj} among others. Typical function spaces for this purpose are modulation and Wiener amalgam spaces, which may serve as space of symbols as well as background where to investigate boundedness and related properties (algebras for composition, sparsity, diagonalization, etc.). In the same spirit, many results are known on dispersive nonlinear PDEs, in particular on the Schr\"odinger equation. Notably, function spaces of time-frequency analysis enjoy a fruitful balance between nice properties (Banach spaces/algebras, embeddings, decomposition, etc.) and regularity of their members. 

The purpose of this overview is to concisely witness some results of this successful interplay, which has made possible to advance in the quest for a rigorous theory of path integral with remarkable results. In particular, we are going to describe three recent contributions on the topic: 
\begin{enumerate}
	\item convergence of time-slicing approximations in $L^p$ spaces (with loss of derivatives) for $p\ne 2$ - based on \cite{STnicola1 conv lp};
	\item convergence of non-smooth time-slicing approximations inspired by the custom in physics and chemistry - based on \cite{STnt};
	\item pointwise convergence of the integral kernels of the sequential approximations - based on \cite{STnt pointw}.	
\end{enumerate}
First we provide a concise exposition of the two operator-theoretic approaches to path integrals mentioned above. We also collect some preliminary concepts in a separate section for the sake of clarity. 

{\bf{Notation.}} We denote by $\STcS(\STrd)$ the Schwartz space of rapidly decaying smooth functions on $\STrd$ and by $\STcS'(\STrd)$ the space of temperate distributions. We set $\langle x \rangle = (1+|x|)^{1/2}$, $x\in \STrd$. The space of smooth bounded functions on $\STrd$ with bounded derivatives of any order is denoted by $C^{\infty}_b(\STrd)$ (also known as $S^{0}_{0,0}$ in microlocal analysis); it is equipped with the family of seminorms 
\[ \left\Vert f \right\Vert_k = \sup_{|\alpha|\le k} \left\Vert \partial^\alpha f \right\Vert_{L^\infty} <\infty, \quad k \in \STbN_0={0,1,2,\ldots}. \] The conjugate exponent $p'$ of $p \in [1,\infty]$ is defined by $1/p+1/p'=1$. We write $f \lesssim g$ if the underlying inequality holds up to a constant factor $C>0$, that is $f\le Cg$. 

\section{A few facts on modulation spaces}\label{STsec mod sp} In this section we set the function space framework for the rest of the paper. The reader is urged to consult \cite{STgro1 book,STtoft cont 1,STtoft cont 2,STwang} for more details and the proofs of the mentioned properties. 

Modulation spaces were introduced by Feichtinger in the '80s \cite{STfei new segal,STfei modulation 83}. At first, they can be thought of as Besov spaces with cubic geometry, namely characterized by isometric boxes in the frequency domain instead of dyadic annuli. To be precise, fix an integer $d\ge 1$; for any $1\le p,q \le \infty$ and $s \in \STbR$ we set
\begin{equation}\label{STmodsp bes} M^{p,q}_s(\STrd) \coloneqq \left\{ f \in \STcS'(\STrd) \, : \, \left\Vert f \right\Vert_{M^{p,q}} = \left( \sum_{k \in \STzd} \langle k \rangle^{qs}\left\Vert \square_k f \right\Vert_{L^p}^q \right)^{1/q} < \infty \right\}, \end{equation} where the frequency-uniform decomposition operator $\square_k$ is the Fourier multiplier whose symbol is a suitably smoothed version of the characteristic function of the unit cube with center $k \in \STzd$. Trivial modifications are needed to cover the cases $p,q=\infty$; for the unweighted case $s=0$ we write $M^{p,q}(\STrd)$, while the case corresponding to $p=q$ simply becomes $M^p(\STrd)$. We emphasize that, in heuristic terms, the parameter $s \ge 0$ can be interpreted as the degree of fractional differentiability of $f \in M^{p,q}_{s}$.

An equivalent, insightful definition of $M^{p,q}(\STrd)$ is known as the \textit{phase-space representation} in the jargon of coorbit theory, namely it is given in terms of the global decay of the phase-space concentration of a distribution. To be concrete, given  $f \in \STcS'(\STrd)$ and a non-zero Schwartz window function $g \in \STcS(\STrd)$, the \textit{short-time Fourier transform} (STFT) $V_g f$ is defined as a windowed version of the ordinary Fourier transform:
\[ V_gf(x,\xi) = \STcF[f \overline{g(\cdot - x)} ](\xi) = \int_{\STrd} e^{- i \xi\cdot t} f(y) \overline{g(t-x)}dt, \qquad (x,\xi) \in \STrdd, \]
where $\STcF$ denotes the Fourier transform. Roughly speaking, the STFT can be interpreted as the magnitude of a tight frequency band centered at $\xi$ in a short time interval centered at $x$. Therefore $V_gf$ can be thought of as a ``musical score'' of the signal $f$, that is an approximately simultaneous time-frequency representation - a perfect phase-space localization is forbidden by the uncertainty principle. The modulation space $M^{p,q}_s(\STrd)$, $1 \le p,q \le \infty$ can then be equipped with the norm 
\[  \left\Vert f \right\Vert_{M^{p,q}_s} = \left( \int_{\STrd} \left( \int_{\STrd} |V_gf(x,\xi)|^p dx\right)^{q/p} \langle \xi \rangle^{qs} d\xi \right)^{1/q}, \] which is proved to be equivalent to the one introduced in \eqref{STmodsp bes} - we improperly use the same notation. Furthermore, different window functions for the STFT yield equivalent norms on $M^{p,q}_s(\STrd)$. We emphasize that several well-known spaces are related with modulation spaces: for instance,
\begin{enumerate}[label=(\roman*)]
	\item $M^2(\STrd)$ coincides with the Hilbert space $L^2(\STrd)$;
	\item $M^2_s(\STrd)$ coincides with the standard $L^2$-based Sobolev space $H^s(\STrd)$;
	\item continuous embeddings with Lebesgue spaces hold:
	\[ M^{p,1}(\STrd) \hookrightarrow L^p(\STrd) \hookrightarrow M^{p,\infty}(\STrd).\] 
\end{enumerate}

A third perspective on modulation spaces is provided by inspecting the definition of the STFT: it may be thought of as a continuous expansion of the function $f$ with respect to the uncountable system $\{\pi(z)g : z=(x,\xi) \in \STrdd \}$, where we introduced the \textit{time-frequency shift operator}
\[ \pi(z) = \pi(x,\xi) = M_{\xi}T_x, \quad M_{\xi}g(t)=e^{i\xi \cdot t}g(t), \quad T_xg(t) = g(t-x). \] Notice that $\pi(z)g$ is a wave packet highly concentrated near $z$ in phase space. In short, we have $V_gf(x,\xi) = \langle f,\pi(x,\xi)g \rangle$ in the sense of the (extension to the duality $\STcS'-\STcS$ of the) inner product on $L^2$. This remark can be made completely rigorous in the context of \textit{frame theory}, leading to discrete time-frequency representations. In particular, given a non-zero window function $g \in L^2(\STrd)$ and a subset $\Lambda \subset \STrdd$, we call \textit{Gabor system} the collection of the time-frequency shifts of $g$ along $\Lambda$, namely $ \STcG(g,\Lambda) = \{ \pi(z)g \, : \, z \in \Lambda\}$. For the sake of concreteness one may consider regular lattices such as $\Lambda = \alpha \STbZ \times \beta \STbZ = \{ (\alpha k,\beta n) \, : \, k,n \in \STbN \}$, for lattice parameters $\alpha,\beta >0$; in that case we write $\STcG(g,\alpha,\beta)$ for the corresponding Gabor system. 
Recall that a \textit{frame} for a Hilbert space $\STcH$ is a sequence $\{x_j\}_{j\in J} \subset \STcH$ such that for all $x \in \STcH$ \[ A\left\Vert x \right\Vert_{\STcH}^2 \le \sum_{j\in J} |\langle x,x_j\rangle|^2 \le B\left\Vert x \right\Vert_{\STcH}^2, \] for some universal constants $A,B>0$ (frame bounds). In a nutshell, the paradigm of frame theory consists in the following steps: decompose a vector $x$ along the frame; investigate how operators act on those elementary pieces; reconstruct the processed vector. The entire process is encoded by the \textit{frame operator}
\[ S \, : \, \STcH \ni x \mapsto \sum_{j\in J}\langle x,x_j\rangle x_j \in \STcH. \]
If a Gabor system $\STcG(g,\Lambda)$ is a frame for $L^2(\STrd)$ it is called \textit{Gabor frame}. Notice that the Gabor frame operator then reads  $Sf = \sum_{z\in \Lambda} V_gf(z)\pi(z)g$. 
A remarkable result is that, for a window function $g \in M^1(\STrd)$ such that $\STcG(g,\alpha,\beta)$ is a frame for $L^2(\STrd)$, an equivalent discrete norm for $M^{p,q}_s(\STrd)$ is given by 
\[ \left\Vert f \right\Vert_{M^{p,q}_s} = \left( \sum_{n\in \STzd} \left( \sum_{k \in \STzd} |V_g f(\alpha k, \beta n)|^p \right)^{q/p} \langle \beta n \rangle^{qs} \right)^{1/q}. \]

\section{Two rigorous approaches to path integrals}\label{STsec rig appr}
We now briefly outline the main features of a pair of mathematical schemes which are in fact two faces of the same philosophy. While in the literature one may easily notice that different names are interchangeably used for them, we consider the classification below for the sake of clarity. 
\subsection{The sequential approach}\label{STsec seq}
The so-called \emph{sequential approach} to path integrals was first introduced by Nelson in \cite{STnelson} and relies on two basic results. 
First, recall that the free evolution operator for the Schr\"odinger equation $U_0(t)=e^{-\SThbi tH_{0}}$, $H_0=-\hbar^2\triangle/2$, is a Fourier multiplier; routine computation yields the following integral representation \cite[Sec. IX.7]{STreed simon 2}:
\begin{equation} \label{STfree prop int}
e^{-\SThbi tH_{0}}\varphi\left(x\right)=\frac{1}{\left(2\pi it\hbar \right)^{d/2}}\int_{\mathbb{R}^{d}}\exp\left(\SThbi\frac{\left|x-y\right|^{2}}{2t}\right)\varphi(y)dy,\qquad\varphi\in\mathcal{S}(\mathbb{R}^{d}).
\end{equation}
Notice that the phase factor in the integral actually coincides with the action functional evaluated along the line $\gamma_{\mathrm{cl}}(\tau) = y + (x-y)\tau/t$, namely the classical trajectory of a free particle moving from position $y$ at time $\tau=0$ to position $x$ at time $\tau = t$ in the absence of external forces.

Next, we need a result from the theory of operator semigroups. Provided that suitable conditions on the domain of $H_0$ and on the potential $V$ are satisfied (see below), the \textit{Trotter product formula} holds for the propagator generated by $H=H_{0}+V$:
\[
e^{-\SThbi t\left(H_{0}+V\right)}=\lim_{n\rightarrow\infty}\left(e^{-\frac{i}{\hbar}  \frac{t}{n}H_{0}}e^{-\SThbi\frac{t}{n}V}\right)^{n},
\]
where the limit is intended in the strong topology of operators in $L^{2}(\mathbb{R}^{d})$. Combining these two ingredients yields the following representation of the complete propagator $e^{-\SThbi tH}$ as limit of integral operators (cf.\ \cite[Thm.\ X.66]{STreed simon 2}):
\begin{equation} 
e^{-\SThbi t\left(H_{0}+V\right)}\phi(x)=\lim_{n\rightarrow\infty}\left(2\pi \hbar i\frac{t}{n}\right)^{-\frac{nd}{2}}\int_{\mathbb{R}^{nd}}e^{ \SThbi S_{n}\left(t;x_{0},\ldots,x_{n-1},x\right)}\varphi\left(x_{0}\right)dx_{0}\ldots dx_{n-1},\label{STprop ts limit}
\end{equation}
where we set
\[
S_{n}\left(t;x_{0},\ldots,x_{n-1},x\right)=\sum_{k=1}^{n}\frac{t}{n}\left[\frac{1}{2}\left(\frac{\left|x_{k}-x_{k-1}\right|}{t/n}\right)^{2}-V\left(x_{k}\right)\right], \quad x_0=y, \,x_{n}=x.
\]
With the aim of understanding the role of   $S_{n}\left(t;x_{0},\ldots,x_{n}\right)$,
consider the following argument. Given the points $x_{0},\ldots,x_{n-1},x\in\mathbb{R}^{d}$,
let $\overline{\gamma}$ be
the polygonal path (broken line) through the vertices $x_{k}=\overline{\gamma}\left(kt/n\right)$,
$k=0,\ldots,n$, $x_n=x$, parametrized as 
\begin{equation}\label{STbroken}
\overline{\gamma}\left(\tau\right)=x_{k}+\frac{x_{k+1}-x_{k}}{t/n}\left(\tau-k\frac{t}{n}\right),\qquad\tau\in\left[k\frac{t}{n},\left(k+1\right)\frac{t}{n}\right],\qquad k=0,\ldots,n-1.
\end{equation}
Hence $\overline{\gamma}$ prescribes a classical motion with constant
velocity along each segment. The action for this path is thus given
by 
\[
S\left[\overline{\gamma}\right]=\sum_{k=1}^{n}\frac{1}{2}\frac{t}{n}\left(\frac{\left|x_{k}-x_{k-1}\right|}{t/n}\right)^{2}-\int_{0}^{t} V(\overline{\gamma}(\tau))d\tau.
\]
According to Feynman's interpretation formula \eqref{STprop ts limit} can be thought of as an integral over all polygonal paths, where $S_{n}\left(x_{0},\ldots,x_{n},t\right)$ is a finite-dimensional approximation of the action functional evaluated on them. The limiting behaviour for $n\rightarrow\infty$ is now intuitively clear:
the set of polygonal paths becomes the set of all paths and in some sense we recover \eqref{STker path int}. We remark that the custom in Physics community after Feynman is exactly to employ the suggestive formula \eqref{STker path int} as a placeholder for \eqref{STprop ts limit} and the related arguments - see for instance \cite{STgs,STkleinert}. 

For what concerns the assumptions on the potential perturbation $V$ under which the Trotter product formula holds, a standard result shows that it is enough to choose $V$ in such a way that $H_0 + V$ is essentially self-adjoint on $D=D(H_0)\cap D(V)$ in $L^2(\STrd)$, cf.  for instance \cite[Thm. VIII.31]{STreed simon 1}. The power of Nelson's perturbative approach is that one can cover wide classes of wild potentials, such as \textit{Kato potentials}, including finite sums of real-valued functions in $L^p(\STrd)$ with $2p>d$ and $p \ge 2$ \cite[Thm. 8]{STnelson}. 

\subsection{The time-slicing approximation}\label{STsec time slic} We now consider another scheme that could be informally called ``the Japanese way'' to rigorous path integrals, since the leading players in its construction were Fujiwara and Kumano-go, with further developments by Ichinose and Tsuchida. The main references for this approach are the papers \cite{STfujiwara1 fund sol,STfujiwara2 duke,STichinose1,STichinose2,STkumanogo0,STkumanogo1,STkumanogo3} and the monograph \cite{STfujiwara5 book}, to which the reader is referred for further details. 

Let us briefly reconsider equation \eqref{STprop ts limit} and its interpretation in terms of finite-dimensional approximations along broken lines; a similar result can be achieved without recourse to the Trotter formula as detailed below. First, let us specify the class of potentials involved in this approach. 
 
{\bf Assumption (A).} {\it The potential $V:\STbR\times \STrd \to \STbR$ satisfies $\partial^\alpha_x V\in C^0(\STbR\times \STrd)$ for any $\alpha \in \STbN_0^d$ and 
	\[ |\partial^\alpha_x V(t,x)| \le C_\alpha, \quad |\alpha|\ge 2, \quad (t,x)\in \STbR\times \STrd \] for suitable constants $C_\alpha >0$.} \\
For this wide class of smooth, time-dependent, at most quadratic potentials Fujiwara showed \cite{STfujiwara1 fund sol,STfujiwara2 duke} that the propagator $U(t,s)$, $0<s<t$, is an \textit{oscillatory integral operator} (for short, OIO) of the form 
\begin{equation}\label{STUts} U(t,s)\varphi(x)=\frac{1}{(2\pi i \hbar (t-s))^{d/2}} \int_{\STrd} e^{\SThbi S(t,s,x,y)} a(\hbar,t,s)(x,y)\varphi(y)dy, \end{equation}
for some amplitude function $a(\hbar,t,s)\in C^{\infty}_b(\STrdd)$. 
In concrete situations, except for a few cases, there is no hope to compute the exact propagator in an explicit, closed form. Due to this difficulty and inspired by the free particle operator \eqref{STfree prop int}, one is lead to consider approximate propagators (\textit{parametrices}), such as
\begin{equation}\label{STE0 def}  E^{(0)}(t,s)\varphi(x) = \frac{1}{(2\pi i \hbar (t-s))^{d/2}} \int_{\STrd} e^{\SThbi S(t,s,x,y)}\varphi(y)dy. \end{equation}
In view of the previous remarks, this operator is supposed to provide a good approximation of the $U(t,s)$ for $t-s$ small enough. The case of a long interval $[s,t]$ can be treated by means of composition of such operators in the spirit of the time slicing method proposed by Feynman: given a subdivision $\Omega = {t_0,\ldots,t_L}$ of the interval $[s,t]$ such that $s=t_0 < t_1 < \ldots < t_L = t$, we define the operator 
\[ E^{(0)}(\Omega,t,s) = E^{(0)}(t_L,t_{L-1})E^{(0)}(t_{L-1},t_{L-2})\cdots E^{(0)}(t_1,t_0), \]
whose integral kernel $e^{(0)}(\Omega,t,s)(x,y)$ can be explicitly computed from \eqref{STE0 def}.
The parametrix $E^{(0)}(\Omega,t,s)$ is then expected to converge (in some sense) to the actual propagator $U(t,s)$ in the limit $ \omega(\Omega) = \max\{t_{j}-t_{j-1}, \, j=1,\ldots,L\} \to 0$. 

We have not specified the path along which the action functional in \eqref{STE0 def} should be evaluated. A standard choice, inspired by the custom in physics after Feynman \cite{STfeyn2 hibbs}, is the broken line approximation introduced above in \eqref{STbroken}, namely \[ \gamma(\tau) = x_j + \frac{x_{j+1}-x_j}{t_{j+1}-t_j}(\tau-t_j),\quad \tau \in [t_j,t_{j+1}],\quad j=0,\ldots,L. \]
A quite complete theory of path integration in this context has been developed by Kumano-go \cite{STkumanogo0}. In fact, the time-slicing approximation shows its full power when straight lines are replaced by \textit{classical paths}. To be precise, under the  assumptions on the potential detailed above, a short-time analysis of the Schr\"odinger flow reveals that there exists $\delta >0$ such that for $0<|t-s|\le \delta$ and any $x,y \in \STrd$ there exists a unique solution $\gamma$ of the classical equation of motion $\ddot{\gamma}(\tau) = - \nabla V(\tau,\gamma(\tau))$ satisfying the boundary conditions $\gamma(s)=y$, $\gamma(t)=x$. In particular, this can be adapted to the subdivision $\Omega$ by making the separation small enough, namely $\omega(\Omega) \le \delta$. A detailed analysis can be found in \cite[Chap. 2]{STfujiwara5 book}.

Among the large number of results proved in this context we mention two milestones from forerunner papers by Fujiwara. In \cite{STfujiwara1 fund sol} he proved convergence of $E^{(0)}(\Omega,t,s)$ to $U(t,s)$ in the norm operator topology in $\STcB(L^2(\STrd))$ - the space of bounded operators in $L^2$. Under the same hypotheses convergence at the level of integral kernels in a very strong topology was proved in \cite{STfujiwara2 duke}. It should be emphasized that the aforementioned results are given for the higher order parametrices $E^{(N)}(t,s)$, $N\in \STbN_0$, also known as \textit{Birkhoff-Maslov parametrices} \cite{STbirkh,STmaslov} and defined by 
\begin{equation}\label{STEN def fuji}  E^{(N)}(t,s)\varphi(x) = \frac{1}{(2\pi i \hbar (t-s))^{d/2}} \int_{\STrd} e^{\SThbi S(t,s,x,y)}a^{(N)}(\hbar,t,s)(x,y)\varphi(y)dy, \end{equation} where $a^{(N)}(\hbar,t,s)(x,y) = \sum_{j=1}^{N} (\SThbi)^{1-j}a_j(t,s)(x,y)$ for suitable functions $a_j(t,s) \in C^{\infty}_b(\STrdd)$ for $t-s\le \delta$, with $a_0(t,s)\equiv 1$. 
We remark that $E^{(N)}(t,s)$ are parametrices in the sense that they satisfy
\begin{equation}\label{STGN def} (i\hbar \partial_t + \hbar^2 \triangle/2 - V(t,x))E^{(N)}\psi = G^{(N)}(t,s)\psi, \end{equation} where $G^{(N)}(t,s)$ has the form in \eqref{STEN def fuji} but $a^{(N)}$ is replaced by the amplitude function $g^{(N)}(\hbar,t,s)(x,y)$ which satisfies $\left\Vert g^{(N)}(\hbar,t,s) \right\Vert_m \le C_m \hbar^{N+1}|t-s|^{N+1}$, $m \in \STbN_0$. 
As before, the case of a long interval $[s,t]$ can be treated by means of composition over a sufficiently fine subdivision $\Omega = {t_0,\ldots,t_L}$ of the interval $[s,t]$ such that $s=t_0 < t_1 < \ldots < t_L = t$, namely
\begin{equation} \label{STEN long}
E^{(N)}(\Omega,t,s) = E^{(N)}(t_L,t_{L-1})E^{(N)}(t_{L-1},t_{L-2})\cdots E^{(N)}(t_1,t_0).
\end{equation}
The core results of the $L^2$ theory for the time slicing approximation read as follows.
\begin{theorem}\label{STmaint L2}
	Let the potential $V$ satisfy Assumption {\rm{(A)}} and fix $T>0$. For $0<t-s\le T$ and any subdivision $\Omega$ of the interval $[s,t]$ such that $\omega(\Omega)\le \delta$, the following claims hold. 
	\begin{enumerate} 
		\item There exists a constant $C=C(N,T)>0$ such that
		\begin{equation}\label{STmain est l2} \left\Vert E^{(N)}(\Omega,t,s)-U(t,s)\right\Vert_{L^2\to L^2} \le C\hbar^{N} \omega(\Omega)^{N+1} (t-s),\quad N \in \STbN_0. \end{equation} 
		\item We have (cf. \eqref{STUts}) \[ \lim_{\omega(\Omega)\to 0} a^{(N)}(\Omega,\hbar,t,s) = a(\hbar,t,s) \quad \text{in }\quad C^{\infty}_b(\STrdd). \] Precisely, there exists $C=C(m,N,T)>0$ such that 
		\[ \left\Vert a(\hbar,t,s)-a^{(N)}(\Omega,\hbar,t,s) \right\Vert_m \le C\hbar^N \omega(\Omega)^{N+1}(t-s), \quad m,N \in \STbN_0. \]
	\end{enumerate}
\end{theorem}

The proof of these results ultimately relies on fine analysis of OIOs. The underlying overall strategy can be condensed as follows:
\begin{enumerate}
	\item prove that ``time slicing approximation is an oscillatory integral'' (cf. \cite{STfujiwara5 book}), i.e., that the operators arising from \eqref{STEN def fuji} are indeed OIOs under suitable assumptions;
	\item derive precise estimates for the operator norm of such OIOs;
	\item employ the algebra property of $\STcB(L^2)$ in order to deal with composition in \eqref{STEN long}.  
\end{enumerate}
With reference to the last item, we mention that an aspect to be considered is that composition of OIOs results in an OIO only for short times, due to the occurrence of caustics, and in general one should not expect smoothing effects for long times. 

For the sake of completeness we also mention that Nicola showed in \cite{STnicola2 ks} how parts of the conclusions in Theorem \ref{STmaint L2} still hold under weaker regularity assumptions for the potential. Assumption (A) is now replaced by the following one. \\
{\bf Assumption (A').}	{\it The potential $V:\STbR\times \STrd \to \STbR$ belongs to $L^1_{\mathrm{loc}}(\STbR \times \STrd)$ and for almost every $t \in \STbR$ and $|\alpha|\le 2$ the derivatives $\partial_x^\alpha V(t,x)$ exist and are continuous with respect to $x$. Furthermore
	\[ \partial_x^\alpha V(t,x) \in L^{\infty}(\STbR; H^{d+1}_{ul}(\STrd)), \quad |\alpha|=2, \]
	where $H^n_{ul}(\STrd)$, $n\in \STbN$, is the Kato-Sobolev space (also known as uniformly local Sobolev space) of functions $f \in L_{\mathrm{loc}}^1(\STrd)$ satisfying $\left\Vert f \right\Vert_{H^n_{ul}} = \sup_B \left\Vert f \right\Vert_{H^n(B)} < \infty$, the supremum being computed on all open balls $B\subset \STrd$ of radius 1.}
\\
\begin{theorem}[{\cite[Thm. 1.1]{STnicola2 ks}}]
	Let the potential $V$ satisfy Assumption {\rm (A')}. For any $T>0$ there exists $C=C(T)>0$ such that for any $0<t-s\le T$ and any subdivision $\Omega$ of the interval $[s,t]$ with $\omega(\Omega) \le \delta$ and $0<\hbar \le 1$, 
	\[ \left\Vert E^{(0)}(\Omega,t,s)-U(t,s)\right\Vert_{L^2\to L^2} \le C \omega(\Omega) (t-s). \]
\end{theorem} 

\section{Beyond the $L^2$ theory via Gabor analysis} 

In view of the results recalled above it seems that the analysis of convergence of time slicing approximations of path integrals can be suitably conducted at the level of operators on $L^2(\STrd)$, i.e. in the space $\STcB(L^2(\STrd))$ (usually) equipped with the norm operator topology.  

It is then natural to wonder whether there exists an $L^p$ analogue of Theorem \ref{STmaint L2} with $p\ne 2$. We cannot expect a naive transposition of the claim for several reasons. First of all, notice that the Schr\"odinger propagator is not even bounded on $L^p(\STrd)$ for $p\ne 2$. The parabolic geometry of its characteristic manifold implies that a peculiar loss of derivative, ultimately due to dispersion, occurs \cite{STbrenner,STmiyachi}:
\[ \left\Vert e^{i\hbar \triangle} f \right\Vert_{L^p} \le C \left\Vert (1-\hbar \triangle)^{k/2}f \right\Vert_{L^p}, \quad k=2d|1/2-1/p|,\quad 1<p<\infty. \] 

On the basis of this observation one is lead to consider the following scale of semiclassical $L^p$-based Sobolev spaces: for $1<p<\infty$ and $k\in \STbR$ define
\[ \STtlpk(\STrd) = \{ f\in S'(\STrd) \,:\, \left\Vert f \right\Vert_{\STtlpk} = \left\Vert (1-\hbar\triangle)^{k/2}f \right\Vert_{L^p} < \infty \}. \]
We set $L^p_k (\STrd) = \STtlpk(\STrd)$ in the case where $\hbar = 1$. 
This is indeed a suitable setting for the analysis of Schr\"odinger operators, in particular for Fourier integral operators arising as Schr\"odinger propagators associated with quadratic Hamiltonians, cf. \cite{STdan}. 

We are also confronted with another issue: the space of bounded operators $\STtlpk \to L^p$ (or viceversa) is clearly not an algebra under composition. This is a major obstacle for a proficient time slicing approximation, having in mind the construction of the parametrices $E^{(N)}(\Omega,t,s)$ in \eqref{STEN long} and the role of this feature in the $L^2$ setting. 

A possible solution comes from time-frequency analysis, since all these issues become manageable as soon as one transfers the problem to the phase space setting. The first key results in this context are due to Nicola \cite{STnicola1 conv lp} and read as follows - the notation has been introduced in the previous section.

\begin{theorem}[{\cite[Thm. 1.1]{STnicola1 conv lp}}]\label{STmaint lp}
	Assume the condition in Assumption {\rm{(A)}} and let $1<p<\infty$, $k = 2d|1/2-1/p|$. 
	\begin{enumerate}
		\item For any $T>0$ there exists a constant $C=C(T)>0$ such that for all $f \in \STcS(\STrd)$, $|t-s|\le T$ and $0<\hbar \le 1$:
		\[ \left\Vert U(t,s)f \right\Vert_{L^p} \le C \left\Vert f \right\Vert_{\STtlpk}, \quad 1 < p \le 2, \]
		\[ \left\Vert U(t,s)f \right\Vert_{\tilde{L}^p_{-k}} \le C \left\Vert f \right\Vert_{L^p}, \quad 2 \le p < \infty.  \]
		\item For any $T>0$ and $N\in \STbN_0$ there exists a constant $C=C(T)>0$ such that for $0<t-s\le T$ and any subdivision $\Omega$ of the interval $[s,t]$ with $\omega(\Omega)\le \delta$, $f\in \STcS(\STrd)$ and $0< \hbar \le 1$:
		\[ \left\Vert \left( E^{(N)}(\Omega,t,s)-U(t,s)\right) f \right\Vert_{L^p} \le C\hbar^N \omega(\Omega)^{N+1}(t-s)\left\Vert f \right\Vert_{\STtlpk}, \quad 1< p \le 2, \]
		\[ \left\Vert \left( E^{(N)}(\Omega,t,s)-U(t,s)\right) f \right\Vert_{\tilde{L}^p_{-k}} \le C\hbar^N \omega(\Omega)^{N+1}(t-s)\left\Vert f \right\Vert_{L^p}, \quad 2 \le p <\infty. \]
	\end{enumerate}
\end{theorem}

For the sake of clarity we organize the discussion of the relevant aspects in separate sections. 

\subsection{The role of modulation spaces} A phase space perspective can be embraced by recasting the problem in terms of modulation spaces. First of all, notice that the characterizing frequency-uniform decomposition in \eqref{STmodsp bes} is particularly well-suited for the analysis of the Schr\"odinger propagator, see \cite{STwang} for a detailed account. The localized operator $\square_k U_0(t)$ is indeed uniformly bounded on $L^p(\STrd)$, namely  
\[ \left\Vert \square_k U_0(t)f \right\Vert_{L^p} \lesssim (1+|t|)^{d|1/2-1/p|}\left\Vert \square_k f\right\Vert_{L^p}.  \]
Moreover it does satisfy the following $L^p - L^{p'}$ estimate:
\[ \left\Vert \square_k U_0(t)f \right\Vert_{L^p} \lesssim (1+|t|)^{-d(1/2-1/p)}\left\Vert \square_k f\right\Vert_{L^{p'}}, \quad p\ge 2.  \] This is a remarkable improvement with respect to the corresponding dyadic estimate\footnote{We set $\mathscr{Q}_k$ for the Fourier multiplier whose symbol is a suitably smoothed version of the characteristic function of the dyadic annulus $R_k = \{\xi \in \STrd \, : \, 2^{k-1} \le |\xi| < 2^k \}$, $k \in \STbN$; we also define $R_0$ to be the unit ball.}
\[ \left\Vert \mathscr{Q}_k U_0(t)f \right\Vert_{L^p} \lesssim |t|^{-d(1/2-1/p)}\left\Vert \mathscr{Q}_k f\right\Vert_{L^{p'}}, \quad p\ge 2, \] showing a singularity at $t=0$.   

The phase space lifting strategy in \cite{STnicola1 conv lp} crucially relies on the following non-trivial embeddings relating modulation and Sobolev spaces. 
\begin{theorem}[{\cite{STks}}]\label{STthm mod emb}
	Let $1 < p < \infty$ and $k= 2d|1/2-1/p|$. The following embeddings hold:
	\[ L^p_k(\STrd) \hookrightarrow M^p(\STrd)  \hookrightarrow L^p(\STrd), \quad 1 < p \le 2, \]
	\[ L^p(\STrd) \hookrightarrow M^p(\STrd)  \hookrightarrow L^p_{-k}(\STrd), \quad 2 \le p <\infty. \] 
\end{theorem}
As a consequence we get that a bounded linear operator $T\in \STcB(M^p(\STrd))$ extends to a bounded operator $T: L^p_k(\STrd) \to L^p(\STrd)$ in the case where $1<p \le 2$, $k=2d|1/2-1/p|$, that is \[ \left\Vert T \right\Vert_{L^p_k \to L^p} \lesssim \left\Vert T \right\Vert_{M^p \to M^p}. \] Similar arguments apply to the case $p\ge 2$. It is interesting to remark that in the case of the free propagator $U_0(t)$ the estimate arising from the $M^p$ operator norm is sharp, that is 
\[ \left\Vert U_0(t) \right\Vert_{L^p_k \to L^p} \lesssim \left\Vert U_0(t) \right\Vert_{M^p \to M^p} \simeq (1+|t|)^{d|1/2-1/p|}, \] where $1<p<\infty$ (the case $p\ge 2$ follows by duality arguments) and $k$ is as before. 
It is also worth mentioning that moving to phase space has another advantage. Establishing endpoint continuity results for integral operators for $p\ne 2$ requires delicate, highly non-trivial arguments usually involving Hardy-type spaces, cf. \cite{STsss,STstein}; these technical aspects are now hidden behind the embeddings in Theorem \ref{STthm mod emb}.

\subsection{Sparse operators on phase space} From the perspective of phase space analysis the next step should be to study how operators transform Gabor wave packets on phase space. A natural object to investigate this feature is the \textit{Gabor matrix} of an operator $T$, namely 
\begin{equation}\label{STdef gabor matrix} \STcM(T,g,z,w) = |\langle T \pi(z)g,\pi(w)g \rangle|, \quad g\in \STcS(\STrd)\setminus\{0\}, \quad z,w \in \STrdd. \end{equation}  We are going to introduce a class of operators characterized by the sparsity of their Gabor matrix. First, we say that a map $\chi : \STrdd \to \STrdd$ is a \textit{tame canonical transformation} if it is a smooth, invertible symplectomorphism such that
\[ |\partial^\alpha_y \partial^\beta_\eta \chi(y,\eta)| \le C_{\alpha,\beta}, \quad |\alpha|+|\beta|\ge 1, \quad y,\eta \in \STrd. \] 
\begin{definition} Let $\chi$ be a tame canonical transformation. We define $FIO(\chi)$ to be the collection of operators $T:\STcS(\STrd)\to \STcS'(\STrd)$ such that for some (hence any) $g \in \STcS(\STrd)\setminus \{0\}$ and any $s\ge 0$  
	\[ \left\Vert T \right\Vert_{s,\chi} = \sup_{z,w \in \STrdd}\langle w-\chi(z)\rangle^{s}\STcM(T,g,z,w) < \infty. \]
$\{ \left\Vert T \right\Vert_{s,\chi}, s\ge 0\}$ is a family of seminorms for $FIO(\chi)$.
\end{definition} 
In heuristic terms this definition encodes the property of operators $T \in FIO(\chi)$ of being concentrated along the graph of $\chi$ in phase space. By an abuse of language we could say that $T$ is \textit{almost diagonalized by Gabor systems} - see also Section \ref{STweyl sec} for further details. 

For future purposes we define a semiclassical version $FIO_{\hbar}(\chi)$ of $FIO(\chi)$ as the space of operators $T:\STcS(\STrd)\to \STcS'(\STrd)$ such that $D_{\hbar^{-1/2}}TD_{\hbar^{1/2}}\in FIO(\chi)$, where we introduced the canonical dilation operator $D_{\hbar^{1/2}}f(t)= \hbar^{-d/4}f(\hbar^{-1/2}t)$. $FIO_{\hbar}(\chi)$ is equipped with the family of seminorms $\left\Vert T \right\Vert_{s,\chi}^{\hbar} = \left\Vert D_{\hbar^{-1/2}}TD_{\hbar^{1/2}} \right\Vert_{s,\chi}$, $s\ge 0$. 

The key properties of the class $FIO(\chi)$ are summarized in the following result. 
\begin{theorem}\label{STthm fio prop}
	\begin{enumerate}
		\item Any operator $T\in FIO(\chi)$ extends to a bounded operator on $M^p(\STrd)$ for $1\le p \le \infty$ - and in particular on $L^2(\STrd) = M^2(\STrd)$. 
		\item The composition $T^{(1)}T^{(2)}$ of operators $T^{(j)} \in FIO (\chi_j)$, $j=1,2$, belongs to $FIO(\chi_1\circ \chi_2)$. 
		\item Assume the hypotheses and notation introduced in Section \ref{STsec time slic}. For $0<|t-s|\le \delta$ and $a \in C^{\infty}_b(\STrdd)$, the oscillatory integral operator 
		\begin{equation}\label{STOIO FIO} Tf(x) = \frac{1}{(2\pi i (t-s)\hbar)^{d/2}}\int_{\STrd} e^{\SThbi S(t,s,x,y)} a(x,y)f(y)dy \end{equation} belongs to $FIO_{\hbar}(\chi^{\hbar}(t,s))$ for a suitable canonical transformation $\chi^{\hbar}(t,s)$ satisfying $\chi^{\hbar}(t,s)=\chi^{\hbar}(t,s)\circ \chi^{\hbar}(t,s)$. Moreover, for any $m \in \STbN_0$ there exist $m' \in \STbN_0$ and a universal constant $C>0$ such that $\left\Vert T \right\Vert_{m,\chi^{\hbar}(t,s)}^{\hbar} \le C \left\Vert a \right\Vert_{m'}$.
		\item Let $T \in FIO_{\hbar}(\chi)$, $1 <p < \infty$ and $k=2d|1/p-1/2|$. Then $T$ extends to a bounded operator $T:\tilde{L}^p_k(\STrd) \to L^p(\STrd)$ if $1<p\le 2$ and $T:L^p(\STrd)\to \tilde{L}^p_{-k}(\STrd)$ if $2\le p < \infty$. In particular, for $m>2d$ there exists $C>0$ such that 
		\begin{equation}\label{STest T lplpk} \left\Vert T \right\Vert_{\tilde{L}^p_k\to L^p} \le C\left\Vert T \right\Vert_{m,\chi}^{\hbar} \,\, (1<p\le 2), \quad \left\Vert T \right\Vert_{L^p\to\tilde{L}^p_{-k}} \le C\left\Vert T \right\Vert_{m,\chi}^{\hbar} \,\, (2\le p < \infty). \end{equation}
	\end{enumerate}
\end{theorem}

We address the reader to \cite{STCdGN semicl,STnicola1 conv lp} for details and proofs. 

It is worthwhile to compare the quite natural proof of the algebra property with the painful arguments concerning the composition of OIOs in \cite{STfujiwara5 book}. Semiclassical versions of Theorems \ref{STthm mod emb} and \ref{STthm fio prop} are needed (and proved) in \cite{STnicola1 conv lp}; see also \cite{STCdGN semicl}. Nonetheless, the first part of Theorem \ref{STmaint lp} essentially follows by noticing that the short-time propagator $U(t,s)$ in \eqref{STUts} is an OIO as in \eqref{STOIO FIO}, that is $U(t,s) \in FIO_{\hbar}(\chi^\hbar(t,s))$; a careful management of the algebra property of $FIO^{\hbar}(\chi_\hbar(t,s))$ and the estimates \eqref{STest T lplpk} yields the claimed result for $0<t-s<T$. 

\subsection{An unavoidable dichotomy}\label{STsec mehler} The peculiar dichotomy in Theorem \ref{STmaint lp} cannot be avoided. A simple argument shows indeed that these results are completely sharp and characterize all possible $L^p$ estimates for time-slicing approximations. Fix $\hbar =1 $ for simplicity and  consider the standard harmonic oscillator, namely
\[ i\partial_t \psi = -\frac{1}{2}\triangle \psi + \frac{1}{2}|x|^2 \psi. \] The propagator can be explicitly computed, its integral kernel is known as the \textit{Mehler kernel} \cite{STdG symp met,STkapit}: for $k \in \STbZ$,
\begin{equation}\label{STmehler} u_t(x,y) = \begin{cases} c(k) |\sin t|^{-d/2}\exp \left( i \frac{x^2+y^2}{2\tan t} - i \frac{x\cdot y}{\sin t} \right) & (\pi k < t < \pi(k+1)) \\ c'(k)\delta((-1)^k x-y) & (t=k\pi) \end{cases} \end{equation} for suitable phase factors $c(k),c'(k)\in \mathbb{C}$.
Notice then that for $t=\pi/2$ the propagator $U(t)$ coincides with the ordinary Fourier transform up to constant factors. The following result implies the sharpness of Theorem \ref{STmaint lp}. 

\begin{theorem}[{\cite[Prop. 7.1]{STnicola1 conv lp}}] Let $1< p < \infty$ and $k_1,k_2 \in \STbR$. The Fourier transform is bounded $L^p_{k_1}(\STrd) \to L^p_{k_2}(\STrd)$ if and only if 
	\[ k_1 \ge 2d(1/p-1/2), \quad k_2 \le 0 \quad (1 < p \le 2), \]
	\[ k_1 \ge 0, \quad k_2 \le -2d(1/2-1/p) \quad (2\le p < \infty). \]
\end{theorem}

\section{Higher order rough parametrices}\label{STsec rough}
Let us come back to Fujiwara's main result, Theorem \eqref{STmaint L2}, to make some important remarks. First, the occurrence of convergence results at two different levels, a coarser one (parametrices in $\STcB(L^2(\STrd))$) and a finer one (OIO amplitudes in $C^{\infty}_b(\STrdd)$), suggests that the assumptions may be relaxed in order to preserve convergence in operator norm - we will devote the subsequent section to the convergence problem for integral kernels. A first step in this direction is the aforementioned paper \cite{STnicola2 ks} by Nicola, where a delicate analysis of low-regular potentials leads to the desired result. We are now going to consider another class of non-smooth potentials as in \cite{STnt}, inspired again by time-frequency analysis function spaces. \\
{\bf Assumption (B).} {\it $V(t,x)$ is a real-valued function of $(t,x)\in\STbR\times \STrd$ and there exists $N\in \STbN$, $N\ge 1$, such that\footnote{We denote by $C^0_b(\STbR,X)$ the space of continuous and bounded functions $\STbR\to X$.}
	\begin{equation}\label{STassumV}
	\partial_t^k \partial_x^\alpha V \in C^0_b(\STbR,M^{\infty,1}(\STrd)),
	\end{equation}
	for any $k\in\STbN$ and $\alpha\in \STbN^d$ satisfying $2k+|\alpha|\leq 2N.$ } \\
The modulation space $M^{\infty,1}(\STrd)$ is also known as the \textit{Sj\"ostrand class}, since it was first introduced by Sj\"ostrand in \cite{STsjo} as an exotic class of symbols still yielding bounded pseudodifferential operators on $L^2(\STrd)$. It was later established that symbols in this space associate with bounded operators on any modulation space and enjoy a rich operator-algebraic structure \cite{STgro ped,STgro2 sj}. As a rule of thumb, a function in $M^{\infty,1}(\STrd)$ is bounded and continuous on $\STrd$; see Section \ref{STsec pw} for further details on the regularity of these functions.
Roughly speaking, potentials satisfying Assumption {\rm{(B)}} are bounded continuous functions together with a certain number of derivatives. Assumptions in the same spirit, or even stronger, are quite popular in scattering theory \cite{STmelrose}. 

In the second place, the estimate \eqref{STmain est l2} reveals other interesting aspects of the parametrices $E^{(N)}$. In particular, notice that while the approximation power increases with $N$ from the point of view of semiclassical analysis (positive powers of $\hbar$), the rate of convergence with respect to the length of the time interval does not enjoy any improvement. Moreover, sophisticate parametrices like those introduced in \eqref{STEN def fuji} have limited applicability to concrete situations and computational problems since the knowledge of the exact action functional is required, the latter being an intractable problem except for a number of simple systems. These remarks lead one to consider short-time approximations for the action by means of the so-called \textit{midpoint rules}. In short, given the action functional corresponding to the standard Hamiltonian $H(q,p,t)= p^2/2 + V(q)$, that is  
\[ S(t,s,x,y) = \frac{|x-y|^2}{2(t-s)} - \STcV(t,s,x,y), \quad \STcV= \int_{s}^{t}V(\gamma(\tau))d\tau, \]
the latter integral involving paths with $\gamma(s)=y$ and $\gamma(t)=x$, $\STcV$ is replaced with approximate expressions such as
\[ \STcV_1= \frac{V(x)+V(y)}{2}(t-s), \quad \text{or }\quad \STcV_2=V\left( \frac{x+y}{2}\right)(t-s).\] 
A simple test in the case of known models reveals that, in spite of their popularity within the physics literature, these procedures are not sufficiently accurate. For the harmonic oscillator and the corresponding approximate actions $S_1,S_2$ one has indeed
\[ S(t,s,x,y)-S_j(t,s,x,y) = O(t-s), \quad j=1,2. \] 
The quest for a correct short-time approximation was initiated by Makri and Miller \cite{STmakri1,STmakri2,STmakri3}, leading to the rule
\[  \overline{\STcV}(s,x,y)=\int_0^1 V(\tau x + (1-\tau)y,s)d\tau. \] 
This procedure satisfies a correct first-order approximation, i.e. $S(t,s,x,y)-\overline{S}(t,s,x,y) = O((t-s)^2)$ for small $t-s$. 
We refer the interested reader to the aforementioned papers and the recent one \cite{STdg short} by de Gosson. 

Inspired by this discussion and by the current practice in physics and chemistry we consider different time slicing approximation operators than \eqref{STEN def fuji}, namely
\begin{equation}\label{STdef EN rough}
\STten(t,s)\varphi(x) = \frac{1}{(2\pi i \hbar (t-s))^{d/2}} \int_{\STrd} e^{\SThbi S^{(N)}(t,s,x,y)}\varphi(y)dy,
\end{equation}
where the approximate action $S^{(N)}$ is essentially a Taylor-like expansion of the exact action $S$ at $t=s$:
\begin{equation}\label{STdef SN}
S^{(N)}(t,s,x,y)=\frac{|x-y|^2}{2(t-s)}+\sum_{k=1}^N W_k(s,x,y)(t-s)^k.
\end{equation}
The coefficients $W_k(s,x,y)$ are recursively constructed after careful analysis of power series solutions for the modified Hamilton-Jacobi equation
\[ \frac{\partial S}{\partial t}+\frac{1}{2}|\nabla_x S|^2+V(t,x)+\frac{i\hbar d}{2(t-s)} - \frac{i\hbar}{2}\Delta_x S=0. \] 
The last two terms are tailored to enhance the approximating power of $\STten$ as parametrix, as showed below. Nevertheless, the ``counterterm'' is first order in $\hbar$ and identically vanishes in the free particle case ($V=0$). Plus, we remark that $W_1(s,x,y) = \overline{\STcV}(s,x,y)$ as expected.

The main properties of these parametrices are summarized below - proofs can be found in \cite{STnt}. 
\begin{theorem}\label{STprop EN rough} Let $V$ satisfy Assumption {\rm (B)} above and let $t,s,T>0$ be such that $0<t-s\le T\hbar $, $0<\hbar \le 1$. 
	\begin{enumerate}
		\item There exists $C=C(T)>0$ such that $\|\STten(t,s)\|_{L^2\to L^2}\leq C$. Moreover, $\STten(t,s) \to I$ (identity op.) for $t\to s$ in the strong operator topology on $L^2$. 
		\item  $\STten(t,s)$ is a parametrix in the sense that
		\[ \left(i\hbar\partial_t+\frac{1}{2}\hbar^2\Delta-V(t,x)\right) \STten(t,s)=G^{(N)}(t,s),\] 
		\[ G^{(N)}(t,s)f =\frac{1}{(2\pi i (t-s) \hbar)^{d/2}} \int_{\STrd} e^{\tfrac{i}{\hbar}S^{(N)}(t,s,x,y)} g_N(t,s,x,y) f(y)\, dy,\] where the amplitude $g_N$ satisfies 
		\[ 	\left\Vert g_N\left(t,s,\cdot,\cdot \right)\right\Vert_{M^{\infty,1}(\STrdd)}\le C\left(t-s\right)^N, \] for some $C=C(T)>0$. 
		\item There exists a constant $C=C(T)>0$ such that
		\begin{equation}\label{STstimaN+1eq} \Vert \STten(t,s) - U(t,s) \Vert_{L^2\rightarrow L^2} \leq C\hbar^{-1}(t-s)^{N+1}.\end{equation}
	\end{enumerate}
\end{theorem}

These estimates should be compared with those appearing in Section \ref{STsec time slic}. Similarly, given a subdivision $\Omega = {t_0,\ldots,t_L}$ of the interval $[s,t]$ such that $s=t_0 < t_1 < \ldots < t_L = t$, we introduce the long-time composition 
\[ \STten(\Omega,t,s) = \STten(t_L,t_{L-1})\STten(t_{L-1},t_{L-2})\cdots \STten(t_1,t_0), \]
and the main result in \cite{STnt} reads as follows. 

\begin{theorem}[{\cite[Thm. 1]{STnt}}]\label{STmaint L2 rough}
	Let $V$ satisfy Assumption {\rm (B)} above. For any $T>0$ there exists a constant $C=C(T)>0$ such that, for $0<t-s\leq T\hbar $, $0 < \hbar \le 1$, and any sufficiently fine subdivision $\Omega$ of the interval $[s,t]$, we have 
	\begin{equation}\label{STstimadapro}
	\|\STten(\Omega,t,s)-U(t,s)\|_{L^2\to L^2}
	\leq C\omega(\Omega)^{N}.
	\end{equation}
\end{theorem}

The proof of Theorem \ref{STmaint L2 rough} is largely inspired by the proof of Theorem \ref{STmaint L2}. In fact, one can isolate a strategy of general interest which can be applied to suitable operators, cf. \cite[Thm. 10]{STnt}.

\subsection*{The role of $\hbar$} We already remarked that Birkhoff-Maslov parametrices \eqref{STEN def fuji} enjoy several nice properties, one of them being an increasing semiclassical approximation power - the exponent of $\hbar$ in \eqref{STmain est l2} increases with $N$. This is of course related to the construction of the parametrices, relying on piecewise classical paths. This desirable property is lost when one considers rougher parametrices as those in \eqref{STdef EN rough}, where the balance weights in favour of accelerated rate of convergence with respect to time. A cursory glance at the estimates for the operators $\STten$ in Theorem \ref{STprop EN rough} reveals that negative powers of $\hbar$ are involved, making them completely unfit for semiclassical arguments. Nevertheless, one can also notice that all the estimates are uniform in $\hbar$ as soon as time is measured in units of $\hbar$, which is a particularly interesting feature. 

\subsection*{The role of $M^{\infty,1}$} Although being hidden in the details of the proofs, the role of the Sj\"ostrand class $M^{\infty,1}(\STrd)$ is crucial for the results presented insofar. There is in particular a special feature of this space playing a major role in the arguments, namely the fact that it is a commutative Banach algebra under pointwise product. 
In general, precise conditions on $p$, $q$, $s$ and $d$ must hold in order for $M_{s}^{p,q}(\STrd)$ to be a Banach algebra with respect to pointwise multiplication\footnote{To be precise, the result provided here concerns conditions under which the embedding $M_{s}^{p,q}\cdot M_{s}^{p,q}\hookrightarrow M_{s}^{p,q}$ is continuous; this means that the algebra property eventually holds up to a constant. It is well known that one may provide an equivalent norm for which the boundedness estimate holds with $C=1$ (cf.\ \cite[Thm.\ 10.2]{STrudin fa}). This condition will be tacitly assumed whenever concerned with Banach algebras from now on.}. 
\begin{proposition}[{\cite[Thm. 3.5 and Cor. 2.10]{STrs mod}}] \label{STMpqs ban alg}Let $1\le p,q\le\infty$ and $s\in\mathbb{R}$. The following facts are equivalent.\\
	$(i)$ $M_{s}^{p,q}(\mathbb{R}^{d})$ is a Banach algebra for pointwise multiplication. \\ $(ii)$ $M^{p,q}_s(\STrd) \hookrightarrow L^{\infty}(\STrd)$. \\ $(iii)$ Either $s=0$ and $q=1$ or $s>d/q'$. 
\end{proposition}

\section{Pointwise convergence of integral kernels}\label{STsec pw}
A concise way to resume the philosophy behind the operator-theoretic approaches to rigorous path integral discussed in Section \ref{STsec rig appr} could be the following one: design suitable sequences of approximation operators and prove that they are bounded together with their compositions, where the latter should converge to the exact propagator in a suitable topology on $\STcB(L^2(\STrd))$. There are good reasons for not being completely satisfied with this state of affairs. First of all, looking back at Feynman's original paper \cite{STfeyn1 48} and the textbook \cite{STfeyn2 hibbs} one immediately notices that the entire process of defining path integrals in \eqref{STker path int} can be read in terms of a sequence of  integral operators (finite-dimensional approximation operators as in \eqref{STprop ts limit} or \eqref{STEN def fuji}); in particular, Feynman's insight calls for the \textit{pointwise} convergence of their integral kernels to the kernel $u_t$ of the propagator. This remark strongly motivates a focus shift from the operators to their kernels, which may appear as an unaffordable problem in general: approximation operators should be first explicitly characterized as integral operators, at least in the sense of distributions by some version of Schwartz's kernel theorem, then one should determine if the kernels are in fact functions and finally hope for convergence. Both the approximation schemes discussed insofar are well suited for this purpose, since oscillatory integrals are explicitly involved. A clue in this direction, already mentioned at the beginning of the previous section, is that the regularity assumptions in Theorem \ref{STmaint L2} imply short-time convergence in a finer topology at the level of integral kernels. 

The solution of this problem in the framework of the sequential approach as presented in Section \ref{STsec seq} was recently obtained by the author and Nicola in the paper \cite{STnt pointw}, where techniques of time-frequency analysis of functions and operators are heavily used. We need a few preparation in order to state the main result.

\textbf{A word of warning about notation}. \textit{In this section we restore the ``harmonic analysis'' normalization of the Fourier transform with $2\pi$ in the phase factor. This reflects into the definition of Weyl, Wigner and short-time Fourier transforms, in contrast with the ``PDE'' normalization adopted insofar. We are sorry if this choice may cause confusion but the aim is to clean up the relevant formulas from annoying normalization constants. For the same reason we set $\hbar = 1$ from now on. }

\subsection{Weyl operators} \label{STweyl sec} A summary of the fruitful exchange between analysis of pseudodifferential operators and time-frequency analysis is far beyond the purposes of this note. The crucial point of contact is represented by the Wigner distribution
\[ W(f,g)(x,\xi )=\int_{\mathbb{R}^{d}}e^{-2\pi iy\cdot \xi }f\left(x+\frac{y}{2}\right)%
\overline{g\left(x-\frac{y}{2}\right)}\ dy, \quad f,g \in f,g\in\mathcal{S}(\mathbb{R}^{d}), \] 
which is a well-known phase space transform deeply connected with the STFT \cite{STgro1 book,STdG symp met}. We define the Weyl transform $\sigma^{\mathrm{w}}:\mathcal{S}(\mathbb{R}^{d})\rightarrow\mathcal{S}'(\mathbb{R}^{d})$ of the symbol $\sigma\in\mathcal{S}'(\mathbb{R}^{2d})$ by duality as follows:
\begin{equation}
\langle \sigma^{\mathrm{w}}f,g\rangle =\langle \sigma,W(g,f)\rangle ,\qquad f,g\in\mathcal{S}(\mathbb{R}^{d}).\label{STdef wig dual}
\end{equation}

As an elementary example of Weyl operator consider the multiplication by $V(x)$, whose symbol is trivially given by \[ \sigma_V (x,\xi)=V(x)=(V\otimes 1)(x,\xi), \quad (x,\xi)\in \STrdd.\] 

The composition of Weyl transforms induces a bilinear form on symbols, the so-called \emph{twisted/Weyl product}: $\sigma^{\mathrm{w}} \circ \rho^{\mathrm{w}} = (\sigma \# \rho)^{\mathrm{w}}$. 

Explicit formulas for the twisted product of are known (cf. \cite{STwong}) but we are more interested in the algebra structure induced on symbol spaces. It turns out that the Sj\"ostrand $M^{\infty,1}(\mathbb{R}^{2d})$, as well as the family of modulation spaces $M_{s}^{\infty}(\mathbb{R}^{2d})$ with $s>2d$,
enjoy a peculiar double Banach algebra structure:  \begin{itemize}
	\item a commutative one associated with pointwise multiplication, as a consequence of Proposition \ref{STMpqs ban alg};
	\item a non-commutative one associated with the Weyl product of symbols (\cite{STgro3 rze,STsjo}). 
\end{itemize} 

The latter algebra structure has been thoroughly investigated in view of its role in the distinctive sparse behaviour satisfied by pseudodifferential operators with symbols in those spaces - the so called \textit{almost diagonalization property} with respect to time-frequency shifts. Having in mind the Gabor matrix defined in \eqref{STdef gabor matrix}, it can be proved that $\sigma\in M^{\infty}_s(\STrdd)$ if and only if, for some (hence any) $g \in \STcS(\STrd)\setminus\{0\}$,
\[ \STcM(\sigma^{\mathrm{w}}, g, z,w) \le C\langle w-z \rangle^{-s}, \quad z,w\in \STrdd. \] 
Similarly, $\sigma\in M^{\infty,1}(\STrdd) $ if and only if there exists $H\in L^1(\STrdd)$ such that
\[ \STcM(\sigma^{\mathrm{w}},g,z,w) \le H(w-z), \quad z,w\in \STrdd. \]
The consequences of phase space sparsity have been thoroughly studied in the papers \cite{STCGNR fio, STCGNR jmp, STCNR sparsity,STCNT 18,STgro2 sj,STgro3 rze}, mainly in order to extend Sj\"ostrand's theory of Wiener subalgebras of Weyl operators \cite{STsjo} to more general pseudodifferential and Fourier integral operators.

\subsection{Main results} 
In order to state the main results in full generality we need to slightly generalize the free Hamiltonian operator $H_0$ in \eqref{STcauchy schr}. Let $a$ be a real-valued, time-independent, quadratic homogeneous polynomial on $\STrdd$, namely 
\[ a(x,\xi) = \frac{1}{2}xAx + \xi B x + \frac{1}{2}\xi C \xi, \] for some symmetric matrices $A,C\in \STbR^{d\times d}$ and $B\in \STbR^{d\times d}$. Consider then the Weyl quantization of $a$ as above. A classical result in phase space harmonic analysis (see \cite[Sec.\ 15.1.3]{STdG symp met} and also \cite{STCN pot mod, STfolland}) is that the solution of \eqref{STcauchy schr} with $H_0 = a^{\mathrm{w}}$ and $V=0$ is given by 
\[ \psi(t,x)=e^{-it H_0}\varphi(x)=\mu(\STcA_t)\varphi(x), \]
where $\mu(\STcA_t)$ is a \textit{metaplectic operator}, designed as follows. First, the classical phase space flow governed by the Hamilton equations\footnote{The factor $2\pi$ derives from the normalization of the Fourier transform adopted in this section.} \[2\pi \dot{z} = J \nabla_z a(z) = \mathbb{A}, \quad \mathbb{A}= \left(\begin{array}{cc} B & C \\ -A & -B^{\top}\end{array}\right) \in \mathfrak{sp}(d,\STbR), \]
defines a mapping \begin{equation}\label{STblock At} \STbR \ni t  \mapsto \STcA_t = e^{(t/2\pi)\mathbb{A}}= \left(\begin{array}{cc}
A_{t} & B_{t}\\
C_{t} & D_{t}
\end{array}\right) \in \mathrm{Sp}(d,\STbR). \end{equation} In very sloppy terms, the metaplectic map $\mu$ is a double-valued unitary representation of the symplectic group on $L^2(\STrd)$, hence the classical flow $\STcA_t$ is ``lifted'' to a family of unitary operators on $L^2(\STrd)$.   
Under certain circumstances an explicit characterization for $\mu(\STcA_t)$ can be provided: for all $t\in \STbR$ such that $\STcA_t$ is a \textit{free symplectic matrix}, namely such that the upper-right block $B_t$ is invertible, the corresponding metaplectic operator is a \textit{quadratic Fourier transform} - cf. \cite[Sec.\  7.2.2]{STdG symp met}:
	\begin{equation} \label{STmet int formula} \mu(\STcA_t)\phi(x) = c_t \lvert \det B_t\rvert^{-1/2} \int_{\STrd} e^{2\pi i \Phi_t(x,y)} \phi(y) dy, \qquad \phi \in \STcS(\STrd), \end{equation} for some $c_t\in \STbC$, $|c_t|=1$, where 
\begin{equation}\label{STphit}
\Phi_{t}(x,y)=\frac{1}{2}xD_{t}B_{t}^{-1}x-yB_{t}^{-1}x+\frac{1}{2}yB_{t}^{-1}A_{t}y,\qquad x,y\in\mathbb{R}^{d}.
\end{equation}
It is known that $H_{0}$ is a self-adjoint operator on its domain (see \cite{SThormander1 mehler})
\[
D\left(H_{0}\right)=\{ \psi \in L^{2}(\mathbb{R}^{d})\,:\,H_{0}\psi \in L^{2}(\mathbb{R}^{d})\} .
\]
In order for the machinery developed in Section \ref{STsec seq} to hold in the case where $H_0 = a^{\mathrm{w}}$ as above we need to consider a version of Trotter formula which holds for semigroups in more general frameworks (cf. for instance \cite[Cor. 2.7]{STengel}). For our purposes, it is enough to assume that $V$ is a bounded perturbation of $H_0$, namely $V \in \STcB(L^2(\STrd))$; notice that $V \in L^{\infty}(\STrd)$ is then a suitable choice, hence including complex-valued potentials. 

Therefore, under the hypotheses on $H_0$ and $V$ discussed insofar, we have 
\begin{equation}\label{STtrotter maint}
e^{-it\left(H_{0}+V\right)}=\lim_{n\rightarrow\infty}E_{n}(t),\qquad E_{n}(t)=\Big(e^{-i\frac{t}{n}H_{0}}e^{-i\frac{t}{n}V}\Big)^{n},
\end{equation}
with convergence in the strong operator topology in $L^{2}(\mathbb{R}^{d})$. Let us denote by $e_{n,t}(x,y)$ the distribution kernel of $E_{n}(t)$ and by $u_t(x,y)$ that of $U(t)=e^{-it\left(H_{0}+V\right)}$.

We assumed $V \in L^{\infty}(\STrd)$, hence there is some room left for tuning the regularity. A suitable choice is given by the modulation spaces $M^{\infty}_s(\STrd)$, $s>d$, and $M^{\infty,1}(\STrd)$, also in view of the rich algebraic structure already discussed. 

In order to grasp the regularity of functions in this space, recall the definition of the Fourier-Lebesgue space: for $s\in \STbR$ we set
\[
f\in\STcF L_{s}^{1}(\mathbb{R}^{d})\quad\Leftrightarrow\quad\left\Vert f\right\Vert _{\mathcal{F}L_{s}^{1}}=\int_{\mathbb{R}^{d}}\left|\mathcal{F}f\left(\xi\right)\right| \langle \xi \rangle^s d\xi<\infty.
\]
The analytic properties of the involved potentials are briefly collected in the following result. 

\begin{proposition}\label{STprop pot sp}
	\begin{enumerate}
		\item $\cap_{s>0} M_{s}^{\infty}(\mathbb{R}^{d}) = C^{\infty}_b (\mathbb{R}^{d})$.
		\item $M^{\infty}_s(\STrd) \subset M^{\infty,1}(\STrd)$ for $s>d$. 
		\item $M^{\infty,1}(\STrd) \subset (\mathcal{F}L^1)_{\rm loc}(\STrd)\cap L^\infty(\STrd)\subset C^0(\STrd)\cap L^\infty(\STrd)$. 
		\item $(M^{\infty,1})_{\rm loc}(\STrd) = (\mathcal{F}L^1)_{\rm loc}(\STrd)$. 
		\item $\STcF\STcM(\STrd) \subset M^{\infty,1}(\STrd)$, where $\STcF\STcM(\STrd)$ is the space of Fourier transforms of (finite) complex measures on $\STrd$.
	\end{enumerate}
\end{proposition}

Roughly speaking, we have a scale of decreasing regularity spaces. 
\begin{enumerate} \item The first is ``the best of all possible worlds'', that is $C^{\infty}_b(\STrd)$.  
	\item At an intermediate stage we have the scale of modulation spaces $M^{\infty}_s(\STrd)$, $s>d$, populated by bounded continuous functions with decreasing (fractional) regularity as $s\searrow d$.
	\item Finally, we have the maximal space $M^{\infty,1}(\STrd)$ where fractional differentiability is completely lost. Nevertheless it is a space of bounded continuous functions locally enjoying the mild regularity of a function in $\STcF L^1$. 
\end{enumerate} 

We first present our main result for potentials in $M^{\infty}_s(\STrd)$.  

\begin{theorem}
	\label{STmaint minfty}Let  $H_{0}=a^{\mathrm{w}}$
	as discussed above and $V\in M_{s}^{\infty}(\mathbb{R}^{d})$,
	with $s>2d$. Let $\STcA_t$ denote the classical flow associated with $H_0$ as in \eqref{STblock At}. For any $t\in\mathbb{R}$ such that $\STcA_t$ is free, that is $\det B_{t}\neq0$:
	\begin{enumerate}
		\item the distributions $e^{-2\pi i\Phi_{t}}e_{n,t}$, $n\geq 1$,
		and $e^{-2\pi i\Phi_{t}}u_t$
		belong to a bounded subset of $M_{s}^{\infty}(\mathbb{R}^{2d})$;
		\item $e_{n,t}\rightarrow u_t$ in $\left(\mathcal{F}L_{r}^{1}\right)_{\mathrm{loc}}(\mathbb{R}^{2d})$
		for any $0<r<s-2d$, hence uniformly on compact subsets. 
	\end{enumerate}
\end{theorem}

The first part of the claim assures that the kernel convergence problem is well posed in this case - the ``amplitudes'' are bounded continuous functions. The second part precisely characterizes the regularity at which convergence occurs, hence the desired pointwise convergence. 

In view of the first item in Proposition \ref{STprop pot sp} and the related characterization $C^{\infty}(\STrdd) = \bigcap_{r>0} \left( \STcF L^1_r \right)_{\mathrm{loc}}(\STrdd)$, we expect to improve the convergence result in the smooth scenario. 
\begin{corollary}
	\label{STmaint s000} Let  $H_{0}=a^{\mathrm{w}}$
	as discussed above and $V\in C^{\infty}_b(\STrd)$. Let $\STcA_t$ denote the classical flow associated with $H_0$ as in \eqref{STblock At}. For any $t\in\mathbb{R}$ such that $\STcA_t$ is free, that is $\det B_{t}\neq0$:
	\begin{enumerate}
		\item the distributions $e^{-2\pi i\Phi_{t}}e_{n,t}$, $n\geq 1$,
		and $e^{-2\pi i\Phi_{t}}u_t$
		belong to a bounded subset of $C^{\infty}_b(\mathbb{R}^{2d})$;
		\item $e_{n,t}\rightarrow u_t$ in $C^{\infty}(\STrdd)$,
		hence uniformly on compact  subsets together with any derivatives. 
	\end{enumerate}
\end{corollary}

This result should be compared with the second claim in Theorem \ref{STmaint L2} by Fujiwara, which motivated our quest. In spite of the different assumptions and approximation schemes, we stress that our result is almost global in time - more on exceptional times below. 

We conclude with the analogous convergence result for potentials in the Sj\"ostrand class. 

\begin{theorem}
	\label{STmaint sjo} Let  $H_{0}=a^{\mathrm{w}}$
	as discussed above and $V\in M^{\infty,1}(\STrd)$. Let $\STcA_t$ denote the classical flow associated with $H_0$ as in \eqref{STblock At}. For any $t\in\mathbb{R}$ such that $\STcA_t$ is free, that is $\det B_{t}\neq0$:
	\begin{enumerate}
		\item the distributions $e^{-2\pi i\Phi_{t}}e_{n,t}$, $n\geq 1$,
		and $e^{-2\pi i\Phi_{t}}u_t$
		belong to a bounded subset of $M^{\infty,1}(\mathbb{R}^{2d})$;
		\item $e_{n,t}\rightarrow u_t$ in $\left(\mathcal{F}L^{1}\right)_{\mathrm{loc}}(\STrdd)$,
		hence uniformly on compact subsets. 
	\end{enumerate}
\end{theorem}

It seems appropriate to highlight that a typical potential setting in the papers by Albeverio and coauthors \cite{STalbeverio1 inv,STalbeverio2 trace,STalbeverio3 jfa, STalbeverio sugg} and It\^o \cite{STito1,STito2} is ``harmonic oscillator plus a bounded perturbation'', the latter in the form of the Fourier transform of a finite complex measure on $\STrd$. While the cited references rely on completely different mathematical schemes for path integrals (which are manufactured as infinite-dimensional OIOs), in view of the embedding  $\STcF\STcM(\STrd) \subset M^{\infty,1}(\STrd)$ mentioned in Proposition \ref{STprop pot sp} we are able to encompass this class of potentials too.

\subsection{The proof at a glance}
In order to understand why our choice of modulation spaces is suitable for the purpose of pointwise convergence we outline the general strategy of the proof of Theorem \ref{STmaint minfty}. The first step is to express the parametrix $E_n(t)$ in integral form and derive a manageable form of the kernel $e_{n,t}$. The algebra property of $M^{\infty}_s(\STrd)$ will play a crucial role from now on. First, we are able to write 
\[
E_{n}\left(t\right)=\left(e^{-i\frac{t}{n}H_0}e^{-i\frac{t}{n}V}\right)^{n}=\left(\mu\left(\mathcal{A}_{t/n}\right)\left(1+i\frac{t}{n}V_{0}\right)\right)^{n}
\]
for a suitable $V_0=V_{0,n,t} \in M^{\infty}_s(\STrd)$ - see \cite[Lem. 3.2]{STnt pointw}. We now expand the (ordered) product and identify multiplication by $1+itV_0/n$ with a suitable Weyl operator, then the nice intertwining properties of Weyl and metaplectic operators (the so-called \textit{symplectic covariance of Weyl calculus}, cf. \cite[Thm.\ 215]{STdG symp met}) yield
\[ 
E_n(t) = \left[ \prod_{k=1}^{n} \left( I+i\frac{t}{n} \left(\sigma_{V_{0}} \circ \STcA_{-k\frac{t}{n}} \right)^{\mathrm{w}} \right) \right] \mu\left(\STcA_{t/n}\right)^n = a_{n,t}^{\text{w}}\,\mu(\mathcal{A}_{t}),
\] where the first (ordered) product is understood in the Banach algebra $(M^{\infty}_s(\STrd,),\#)$. The symbol of $a_{n,t}^{\mathrm{w}}$ satisfies the estimate $\left\Vert a_{n,t}\right\Vert _{M_{s}^{\infty}}\le e^{C\left(t\right)t}$ for some locally bounded constant $C(t)>0$ independent of $n$. 

Since $\mathcal{A}_{t}$ is a free symplectic matrix, the integral formula \eqref{STmet int formula} holds and with the help of some technical lemmas we are able to precisely characterize the integral kernels $e_{n,t}$ and $u_t$ as temperate distributions. 
The non-trivial step is to prove convergence in $\mathcal{S}'(\mathbb{R}^{2d})$, but it can be handled with Banach algebras techniques and some topological arguments. The assumptions on potentials imply the boundedness of the sequence $\{e_{n,t} \}$ in $M^{\infty}_s(\STrdd)$. Finally, the convergence of $e_{n,t}$ to $u_t$ in $\left(\mathcal{F}L_{r}^{1}\right)_{\mathrm{loc}}(\mathbb{R}^{2d})$, $0<r<s-2d$, essentially follow by dominated convergence arguments. 

The proof of Theorem \ref{STmaint sjo} ultimately moves along the same lines but is more involved, so we will not give the details here. The basic ingredient is a ``high-cut filter decomposition'' of $M^{\infty,1}$, see \cite[Lem. 3.3]{STnt pointw}: a rough function $f\in M^{\infty,1}(\STrd)$ can be splitted as a sum of a very regular part $f_1 \in C^{\infty}_b(\STrd)$ plus an arbitrarily small (in norm) rough remainder $f_2 \in M^{\infty,1}(\STrd)$. Theorem \ref{STmaint sjo} essentially follows as a perturbation of Theorem \ref{STmaint minfty} after a careful management of these remainders.

\subsection{Why exceptional times?} The occurrence of a set of exceptional times in Theorems \ref{STmaint minfty} and \ref{STmaint sjo} is to be expected from a mathematical point of view: it may happen that the integral kernel of the propagator degenerates into a distribution. A well-known example of this behaviour is provided by the harmonic oscillator, already met in Section \ref{STsec mehler}. Mehler's formula \eqref{STmehler} precisely shows the expected degenerate behaviour, which is consistent with the fact that $\STcA_t$ is free if and only if $t \ne k\pi$, $k\in\mathbb{Z}$, since $B_t = (\sin t)I_{d\times d}$ (up to normalization constants). 
The physical interpretation of the exceptional values is not entirely clear at the moment, but milder convergence results in the spirit of the theorems above may be proved to hold also at exceptional times. These and other related issues will be object of forthcoming contributions \cite{STFNT}.


\begin{thebibliography}{99}
	\bibitem{STalbeverio1 inv} S. Albeverio, and R. H\o egh-Krohn. Oscillatory integrals and the method of stationary phase in infinitely many dimensions, with applications to the classical limit of quantum mechanics. I. \textit{Invent. Math.} \textbf{40}(1) (1977), 59--106.
	
	\bibitem{STalbeverio2 trace} S. Albeverio, P. Blanchard, and R. H\o egh-Krohn. Feynman path integrals and the trace formula for the Schr\"odinger operators. \textit{Comm. Math. Phys.} \textbf{83}(1) (1982), 49--76. 
	
	\bibitem{STalbeverio3 jfa} S. Albeverio, and Z. Brze\'zniak. Finite-dimensional approximation approach to oscillatory integrals and stationary phase in infinite dimensions.\textit{ J. Funct. Anal.} \textbf{113}(1) (1993), 177--244.
	
	\bibitem{STalbeverio book} S. Albeverio, R. H\o egh-Krohn, and S. Mazzucchi. {\it Mathematical theory of Feynman path integrals. An Introduction}. Lecture Notes in Mathematics 523. Springer-Verlag, Berlin, 2008.
	
	\bibitem{STalbeverio sugg} S. Albeverio, and S. Mazzucchi. A unified approach to infinite-dimensional integration. \textit{Rev. Math. Phys.} \textbf{28}(2) (2016), 1650005, 43 pp. 
	
	\bibitem{STbirkh} G.D. Birkhoff. Quantum mechanics and asymptotic series. \textit{Bull. Amer. Math. Soc.} \textbf{39} (1933), no. 10, 681--700. 
	
	\bibitem{STbrenner} P. Brenner, V. Thom\'ee, and L.B. Wahlbin. \textit{Besov spaces and applications to difference methods for initial value problems}. Springer-Verlag, Berlin-New York, 1975. 
	
	\bibitem{STfeyn thesis} L. M. Brown (ed.). \textit{Feynman's Thesis. A New Approach to Quantum Theory}. World Scientific, Hackensack, 2005. 
	
	\bibitem{STcameron} R. Cameron. A family of integrals serving to connect the Wiener and Feynman integrals. \textit{J. Math. and Phys.} \textbf{39} (1960), 126--140.
	
	\bibitem{STCdGN semicl} E. Cordero, M. de Gosson, and F. Nicola. Semi-classical time-frequency analysis and applications. \textit{Math. Phys. Anal. Geom.} \textbf{20} (2017), no. 4, Art. 26, 23 pp.
	
	\bibitem{STCGNR fio} E. Cordero, K. Gr\"ochenig, F. Nicola, and L. Rodino. Wiener algebras of Fourier integral operators. \textit{J. Math. Pures Appl.} \textbf{99}(2) (2013), 219--233. 
	
	\bibitem{STCGNR jmp} E. Cordero, K. Gr\"ochenig, F. Nicola, and L. Rodino. Generalized metaplectic operators and the Schr\"odinger equation with a potential in the Sj\"ostrand class. {\it J. Math. Phys.} \textbf{55}(8) (2014), 081506.
	
	\bibitem{STCNR sparsity} E. Cordero, F. Nicola, and L. Rodino. Sparsity of Gabor representation of Schr\"odinger propagators. \textit{Appl. Comput. Harmon. Anal.} \textbf{26} (2009), no. 3, 357--370.
	
	\bibitem{STCN pot mod} E. Cordero and F. Nicola. On the Schr\"odinger equation with potential in modulation spaces. {\it J. Pseudo-Differ. Oper. Appl.} \textbf{5}(3) (2014), 319--341.
	
	\bibitem{STCNT 18} E. Cordero, F. Nicola, and S. I. Trapasso. Almost diagonalization of $\tau $-pseudodifferential operators with symbols in Wiener amalgam and modulation spaces. \textit{J. Fourier Anal. Appl.} \textbf{25} (2019), no. 4, 1927--1957.
	
	\bibitem{STdan} P. D'Ancona, and F. Nicola. Sharp $L^p$ estimates for Schr\"odinger groups. \textit{Rev. Mat. Iberoam.} \textbf{32} (2016), no. 3, 1019--1038.
	
	\bibitem{STdg short} M. de Gosson. Short-time propagators and the Born-Jordan quantization rule. \textit{Entropy} \textbf{20}(11) (2018), 869.
	
	\bibitem{STdG symp met} M. de Gosson. {\it Symplectic Methods in Harmonic Analysis and in Mathematical Physics}. Birkh\"auser/Springer Basel AG, Basel, 2011.
	
	\bibitem{STengel} K.-J. Engel, and R. Nagel. {\it A Short Course on Operator Semigroups}. Springer, New York, 2006.
	
	\bibitem{STFNT} H. G. Feichtinger, F. Nicola and S. I. Trapasso. On exceptional times for pointwise convergence of integral kernels in Feynman-Trotter path integrals. Submitted.
	
	\bibitem{STfei new segal}
	H. G. Feichtinger. On a new {S}egal algebra. {\it Monatsh. Math.} \textbf{92}(4) (1981), 269--289.
	
	\bibitem{STfei modulation 83}
	H. G. Feichtinger. Modulation spaces on locally compact Abelian groups. In \textit{Proc. Int. Conf. Wavelets and Applications}, Allied Publishers, New Delhi, 2003, pp. 99--140. Reprint of 1983.
	
	\bibitem{STfeyn1 48} R. Feynman. Space-time approach to non-relativistic quantum mechanics. {\it Rev. Mod. Phys.} \textbf{20} (1948), 367--387. 
	
	\bibitem{STfeyn 49} R. Feynman. Space-time approach to quantum electrodynamics. \textit{Phys. Rev.} (2) \textbf{76} (1949), 769--789.
	
	\bibitem{STfeyn2 hibbs} R. Feynman, and A.R. Hibbs. {\it Quantum Mechanics and Path Integrals}. Emended Edition. Dover Publications, Mineola, 2005.
	
	\bibitem{STfolland} G.B. Folland. \textit{Harmonic Analysis in Phase Space}. Princeton University Press, Princeton, 1989. 
	
	\bibitem{STfujiwara1 fund sol} D. Fujiwara. A construction of the fundamental solution for the Schr\"odinger equation. {\it J. Anal. Math.} \textbf{35}, 41--96, 1979.
	
	\bibitem{STfujiwara2 duke} D. Fujiwara. Remarks on convergence of some Feynman path integrals. {\it Duke Math. J.} \textbf{47} (1980), 559--600. 
	
	\bibitem{STfujiwara5 book} D. Fujiwara. {\it Rigorous Time Slicing Approach to Feynman Path Integrals}. Springer, Tokyo, 2017.
	
	\bibitem{STgro1 book}  K. Gr\"ochenig. {\it Foundations of Time-frequency Analysis}. Birkh\"auser, Boston, 2001. 
	
	\bibitem{STgro ped} K. Gr\"ochenig. A pedestrian's approach to pseudodifferential operators. In \textit{Harmonic Analysis and Applications}, 139--169, Birkh\"auser, Boston, 2006.
	
	\bibitem{STgro2 sj} K. Gr\"ochenig. Time-frequency analysis of Sj\"ostrand's class. {\it Rev. Mat. Iberoam.} \textbf{22}(2) (2006), 703--724. 
	
	\bibitem{STgro3 rze} K. Gr\"ochenig, and Z. Rzeszotnik. Banach algebras of pseudodifferential operators and their almost diagonalization. \textit{Ann. Inst. Fourier} \textbf{58} (2008), no. 7, 2279--2314. 
	
	\bibitem{STgs} C. Grosche, and F. Steiner. \textit{Handbook of Feynman path integrals}. Springer, Berlin, 1998. 
	
	\bibitem{SThormander1 mehler} L. H\"ormander. Symplectic classification of quadratic forms, and general Mehler formulas. \textit{Math. Z.} \textbf{219} (1995), no. 3, 413--449. 
	
	\bibitem{SThormander2 book 3} L. H\"ormander. \textit{The analysis of linear partial differential operators III. Pseudo-differential operators}. Reprint of the 1994 edition. Classics in Mathematics. Springer, Berlin, 2007.
	
	\bibitem{STichinose1} W. Ichinose. On the formulation of the Feynman path integral through broken line paths. {\it  Comm. Math. Phys} \textbf{189}(1) (1997), 17--33.
	
	\bibitem{STichinose2} W. Ichinose. Convergence of the Feynman path integral in the weighted Sobolev spaces and the representation of correlation functions. {\it J. Math. Soc. Japan} \textbf{55}(4) (2003), 957--983.
	
	\bibitem{STito1} K. It\^o. Wiener integral and Feynman integral. In \textit{Proc. 4th Berkeley Sympos. Math. Statist. and Prob., Vol. II}, 227--238. Univ. California Press, Berkeley, 1961.
	
	\bibitem{STito2} K. It\^o. Generalized uniform complex measures in the Hilbertian metric space with their application to the Feynman integral. In \textit{Proc. 5th Berkeley Sympos. Math. Statist. and Prob., Vol. II }, 145--161. Univ. California Press, Berkeley, 1967.
	
	\bibitem{STkapit} L. Kapitanski, I. Rodnianski, and K. Yajima. On the fundamental solution of a perturbed harmonic oscillator. \textit{Topol. Methods Nonlinear Anal.} \textbf{9}(1) (1997), 77--106.
	
	\bibitem{STkleinert} H. Kleinert. \textit{Path Integrals in Quantum Mechanics, Statistics and Polymer Physics}. World Scientific, Singapore, 1995.
	
	\bibitem{STks} M. Kobayashi and M. Sugimoto. The inclusion relation between Sobolev and modulation spaces. \textit{J. Funct. Anal.} \textbf{260} (2011), no. 11, 3189--3208.
	
	\bibitem{STkumanogo0} N. Kumano-go. Feynman path integrals as analysis on path space by time slicing approximation. {\it Bull. Sci. Math.} \textbf{128}(3) (2004), 197--251.  
	
	\bibitem{STkumanogo1} N. Kumano-go. A construction of the fundamental solution for Schr\"odinger equations. {\it J. Math. Sci. Univ. Tokyo} \textbf{2} (1995), 441--498.
	
	\bibitem{STkumanogo3} N. Kumano-go and D. Fujiwara. Smooth functional derivatives in Feynman path integrals by time slicing approximation. {\it Bull. Sci. Math.} \textbf{129}(1) (2005), 57--79.
	
	\bibitem{STmakri1} N. Makri and W. H. Miller. Correct short time propagator for Feynman path integration by power series expansion in $\delta t$. {\it Chem. Phys. Lett.}, 151, 1--8, 1988.
	
	\bibitem{STmakri2} N. Makri and W. H. Miller. Exponential power series expansion for the quantum time evolution operator. {\it J. Chem. Phys.}, 90(2), 904-911, 1989.
	
	\bibitem{STmakri3} N. Makri. Feynman path integration in quantum dynamics. {\it Comput. Phys. Comm.}, 63(1), 389--414, 1991. 
	
	\bibitem{STmaslov} V.P. Maslov. \textit{Th\'eorie des Perturbations et M\'ethodes Asymptotiques}. (French translation from Russian) Dunod, Paris, 1970.	
	
	\bibitem{STmazzucchi} S. Mazzucchi. {\it Mathematical Feynman Path Integrals and Their Applications}. World Scientific, 2009.
	
	\bibitem{STmelrose} R. Melrose.  \textit{Geometric Scattering Theory}. Cambridge University Press, Cambridge, 1995. 
	
	\bibitem{STmiyachi} A. Miyachi. On some Fourier multipliers for $H^p(\STbR^n)$. \textit{J. Fac. Sci. Univ. Tokyo Sect. IA Math.} \textbf{27} (1980), no. 1, 157--179. 
	
	\bibitem{STnelson} E. Nelson. Feynman integrals and Schr\"odinger equation. \textit{J. Math. Phys.} \textbf{5} (1964), 332--343.	
	
	\bibitem{STnicola1 conv lp} F. Nicola. Convergence in $L^p$ for Feynman path integrals. {\it Adv. Math.} {\bf 294} (2016), 384--409.
	
	\bibitem{STnicola2 ks} F. Nicola. On the time slicing approximation of Feynman path integrals for non-smooth potentials.  {\it J. Anal. Math.} {\bf 137}(2) (2019), 529--558.
	
	\bibitem{STnt} F. Nicola and S. I. Trapasso. Approximation of Feynman path integrals with non-smooth potentials. \textit{J. Math. Phys.} \textbf{60} (2019), 102103. 
	
	\bibitem{STnt pointw} F. Nicola and S. I. Trapasso. On the pointwise convergence of the integral kernels in the Feynman-Trotter formula. \textit{Comm. Math. Phys.} (2019) - DOI: 10.1007/s00220-019-03524-2. 
	
	\bibitem{STreed simon 1} M. Reed and B. Simon. {\it Methods of Modern Mathematical Physics. Vol. I: Functional analysis}. Academic Press, 1981.
	
	\bibitem{STreed simon 2} M. Reed and B. Simon. {\it Methods of Modern Mathematical Physics. Vol. II: Fourier analysis, self-adjointness}. Elsevier, 1975.
	
	\bibitem{STrs mod} M. Reich and W. Sickel. Multiplication and composition in weighted modulation spaces. In \textit{Mathematical analysis, probability and applications  - plenary lectures}, 103--149, Springer Proc. Math. Stat., 177, Springer, 2016. 
	
	\bibitem{STrudin fa} W. Rudin. \textit{Functional analysis}. Second edition. International Series in Pure and Applied Mathematics. McGraw-Hill, New York, 1991.
	
	\bibitem{STsauer} T. Sauer. Remarks on the origin of path integration: Einstein and Feynman. In \textit{Proceedings of the 9th International Conference on Path Integrals: New Trends and Perspectives}, 3--13, World Sci. Publ., Hackensack, NJ, 2008. 
	
	\bibitem{STsss} A. Seeger, C.D. Sogge, and E.M. Stein. Regularity properties of Fourier integral operators. \textit{Ann. of Math.} (2) \textbf{134} (1991), no. 2, 231--251. 
	
	\bibitem{STsjo} J. Sj\"ostrand. An algebra of pseudodifferential operators. {\it Math. Res. Lett.} \textbf{1}(2) (1994), 185--192. 
	
	\bibitem{STstein} E.M. Stein. \textit{Harmonic analysis: real-variable methods, orthogonality, and oscillatory integrals.} Princeton University Press, Princeton, 1993. 
	
	\bibitem{STtoft cont 1} J. Toft. Continuity properties for modulation spaces, with applications to pseudo-differential calculus. I. \textit{J. Funct. Anal.} \textbf{207} (2004), no. 2, 399--429. 
	
	\bibitem{STtoft cont 2} J. Toft. Continuity properties for modulation spaces, with applications to pseudo-differential calculus. II. \textit{Ann. Global Anal. Geom.} \textbf{26} (2004), no. 1, 73--106.
	
	\bibitem{STwang} B. Wang, Z. Huo, C. Hao and Z. Guo. \textit{Harmonic Analysis Method for Nonlinear Evolution Equations}. World Scientific Publishing Co. Pte. Ltd., Hackensack, NJ, 2011. 
	
	\bibitem{STwong} M. W. Wong. \emph{Weyl transforms}. Springer-Verlag, New York, 1998. 
	
\end{thebibliography}
\end{document}